\documentclass[11pt]{article}

\usepackage{amssymb}
\usepackage{amsmath}
\usepackage{epstopdf}

\title{Using Pulsars to Define Space-Time Coordinates}
\author{Bartolom\'e Coll\thanks{Observatoire de Paris -- bartolome.coll@uv.es} \ and Albert Tarantola\thanks{Institut de Physique du Globe de Paris -- albert.tarantola@ipgp.org}}

\begin{document}
\maketitle
 
\begin{abstract}
Fully relativistic coordinates have been proposed for (relativistically) running a ``GPS'' system. These coordinates are the arrival times of the light signals emitted by four ``satellites'' (clocks). Replacing the signals emitted by four controlled clocks by the signals emitted by four pulsars defines a coordinate system with lower accuracy, but valid across the whole Solar System. We here precisely define this new coordinate system, by choosing four particular pulsars and a particular event as the origin of the coordinates.
 
\end{abstract}

We are all familiar with the three coordinates used to represent
the position of points in the laboratory, on the Earth, or in the
sky. Relativity theory has taught us that time is not to be viewed
as flowing independently of space, and physicists routinely use
the notion of event: a point in the four-dimensional continuum
introduced by Poincar\'e (later called space-time by Minkowski),
where an event has four coordinates, say \,$ (x,y,z,t) $\,. An
experimental protocol allowing to attach such four coordinates to
an event, using clocks and light signals, was exposed by Einstein.

One of the major technical advances in the recent years has been
the development of Global Navigation Satellite Systems: GPS in
the USA, Glonass in Russia, and Galileo (under construction) in
the EU. These systems provide the coordinates of any point on
the Earth, as well as an `universal time'. Unfortunately, these
coordinates do not qualify as fully relativistic: in Minkowski
space-time, durations and distances depend on the observer, and
are not universal.
While the Minkowski coordinates \,$ (x,y,z,t) $\, of an event are bona-fide
relativistic coordinates, they are not immediate (using, for
instance, Einstein's protocol, the coordinates are known only
after the delay necessary for light signals to travel to/from
the origin of the system).

Is it possible to define in the space-time an immediate system
of four relativistic coordinates? The answer is yes. Four clocks
in arbitrary motion in space-time, broadcasting their proper time
constitute such a system: the coordinates of a point in space-time
(i.e., of a space-time event) are, by definition, the four times
\,$ (\tau_1,\tau_2,\tau_3,\tau_4) $\, of the four signals converging at that space-time
point, as recorded by a standard receiver. So any observer able
to receive the signals is able to (instantaneously) know its
own coordinates, in fact, its own space-time trajectory, expressed
in these `light-coordinates'\footnote{We say 'light-coordinates because
one would typically use electromagnetic waves to propagate the
signals.}.

But how could these coordinates be used to do space-time geometry,
i.e., how could them be related to the observer's local unit of
time and of distance? We have suggested (Coll, 2000; Coll et al., 2006, 
Tarantola et al.\ 2009), to take
this point of view as the basic paradigm for `running' a Global
Positioning System. Imagine that four `satellites' wander in
space-time, emitting their proper times. If, in addition to
broadcasting its own time signal, each satellite sends an echo
of the signals received from the other three, then, any observer
able to receive the signals of the four satellites knows their
trajectory in this self-consistent coordinate system. Then, this
observer is able to write the components of the space-time metric
tensor at the space-time point where the observer is. In other
words, the observer can use the light-coordinates to build a local
clock and, therefore, a local meter\footnote{This, of course, is only exact if the gravitational field is given. In Tarantola et al.\ (2009) a more general setting is proposed, where the space-time metric itself is evaluated. 
For other details, see http://www.coll.cc.}.

The agency in charge of running a satellite positioning constellation
should only care on the quality of the embarked clocks, and the
quality of the broadcasted signals. Relating these light-coordinates
to any terrestrial coordinate system (either a global geographical
system or a local one) is just an attachment problem, that should not
interfere with the problem of defining the primary coordinate system
itself. Under some simple assumptions about the gravity field around
the Earth, four satellites define a usable coordinate system.
Redundancies created by supplementary satellites should allow to model
the gravity field itself.
Some day, all positioning satellite systems will be run this way.

Besides this, can we today build a relativistic immediate coordinate
system valid in a large domain (larger than our Solar system)?
Yes, if instead of using the signals coming from manufactured and
controlled clocks we use the natural objects that best approach this:
pulsars.

Pulsars are rotating astronomical objects, located in the Galaxy,
that emit quasi-periodic signals. Of particular interest to us are
the millisecond pulsars (with a period of the order of 1 ms).
The main source of variability in the pulsars' signals is the
interstellar medium, that imposes an uncertainty in the measurement
of arrival times of the order of 4 nanoseconds.

We propose to define a Universal (well, say Galactic) system of
space-time coordinates as follows. By convention we select the
four millisecond pulsars 0751+1807 (3.5\,ms), 2322+2057 (4.8\,ms),
0711-6830 (5.5\,ms) and 1518+0205B (7.9\,ms). Their angular distribution
around the Solar system is quite even (they look almost like the
vertices of a tetrahedron seen from its center). We define the origin
\,$ (\tau_1,\tau_2,\tau_3,\tau_4) \, = \, (0,0,0,0) $\, of the space-time coordinates as the event
\emph{ 0H0'0", January 1, 2001}, at the focal point of the Cambridge
radiotelescope (the one that was used for the discovery of the pulsars).
Then, any other space-time event, on Earth, on the Moon, anywhere
in the Solar system or in the solar systems in this part of the
Galaxy, has its own coordinates attributed. With present-day
technology, this locates any event with an accuracy of the order
of 4\,ns, i.e., of the order of one meter. This is not an extremely
precise coordinate system, but it is extremely stable and has a great
domain of validity.

Imagine when spacecrafts that ---like Pioneer Voyager--- leave our
solar system, are equipped with a receiver of the signals from
the four reference pulsars: in their messages to the Earth they
will then include their own space-time coordinates.

If, tomorrow, a human colony is installed in, say, Pluto,
with radiotelescopes able to receive the signals of the four reference
pulsars. It would not be sufficient to e-mail this \emph{arXiv} posting
to the Pluto colony in order for them to be able to use these
coordinates: one vessel must have travelled from the Earth to Pluto,
continuously recording the pulsar signals, so the origin of the
coordinates would have been `transported' to Pluto. Only then, the
Pluto colony would share a common coordinate system with the mainland.

\bigskip

\centerline{---------}

\bigskip

\noindent
We are grateful to F. Biraud for helping
in the selection of the four pulsars).

\section*{References}

\def\bibref{\par\noindent\hangindent=20pt}

\bibref
Coll, B., Proc.\ Spanish Relativistic Meeting, EREs 2000.\\[-12 pt]

\bibref
Coll, B., Ferrando, J.J., Morales, J.A., Positioning with stationary emitters in a two-dimensional space-time, Physical Review D 74, 104003, 2006).

\bibref
Tarantola, A., Klimes, L., Pozo, J.M., and Coll, B., 2009, arXiv: 0905.3798v1, May 2009.

\end{document}